\documentstyle[12pt]{article}  
\begin{document}  \bibliographystyle{unsrt}

\begin{center}
{\Large \bf Wigner's Last Papers on Spacetime Symmetries}~\footnote{based
on a review talk presented at the 4th International Wigner Symposium,
Guadalajara, Mexico (August 1995).}
\\[6mm]

Y. S. Kim\footnote{email: kim@umdhep.umd.edu}   \\
{\it Department of Physics, University of Maryland, \\College Park,
Maryland 20742, U.S.A.}
\end{center}

\vspace{4mm}

\begin{abstract}
Wigner's 1939 paper on representations of the inhomogeneous Lorentz
group is one of the most fundamental papers in physics.  Wigner
maintained his passion for this subject throughout his life.  In this
report, I will review the papers which he published with me on this
subject.  These papers deal with the question of unifying the internal
space-time symmetries of massive and massless particles.
\end{abstract}

\section{Introduction}\label{intro}
I met Eugene Wigner while I was a graduate student at Princeton University
from 1958 to 1961.  I stayed there for one more year as a post-doctoral
fellow before joining the faculty of the University of Maryland in 1962.
My advisor was Sam Treiman, and I wrote my PhD thesis on dispersion
relations.  However, my extra-curricular activity was on Wigner's papers,
particularly on his 1939 paper on representations of the Poincar\'e
group~\cite{wig39}.  It is not uncommon for one's extra-curricular
activity to become his/her life-time job.  Indeed, by 1985, I had
completed the manuscript for the book entitled {\em Theory and
Applications of the Poincar\'e Group}~\cite{knp86} with Marilyn Noz who
has been my closest colleague since 1970.

After writing this book, I approached Wigner again and asked him whether
I could start working on edited volumes of all the papers he had written,
but he had a better idea.  Wigner told me that he was interested in
writing new papers and that he had been looking for a younger person who
could collaborate with him.  This was how I was able to publish seven
papers with him.  Today, I would like to talk about two of those papers.
They constitute a re-interpretation of Wigner's original paper on the
Poincar\'e group.

Why is this paper so important?  Where does it stand in the history of
physics?  From the principles of special relativity, Einstein derived
the relation $E = mc^{2}$ in 1905.  This formula unifies the
momentum-energy relations for both massive and massless particles,
which are $E = p^{2}/2m$ and $E = cp$ respectively.  In his 1939
paper~\cite{wig39}, Wigner observed that relativistic particles have
their internal space-time degrees of freedom.  For example, the spin of
a particle at rest is a manifestation of the three-dimensional rotational
symmetry.  Wigner in his paper formulated space-time symmetries of
relativistic particles in terms of the little groups of the Poincar\'e
group.  In this review talk, I would like to emphasize that Wigner's
little group is a Lorentz-covariant entity and unifies the internal
space-time symmetries of both massive and massless particles, just as
Einstein's $E = mc^{2}$ does for the energy-momentum relation.

On the other hand, Wigner did not reach this conclusion in 1939, and
the above statement is based on many subsequent papers published on
this subject during the period 1939-1990.  In fact, his 1939 paper has a
stormy history.  This paper had been rejected by one of the prestigious
mathematics journals before John von Neumann, then the editor of the
Annals of Mathematics, invited Wigner to submit it to his journal.  It
is not uncommon even these days to hear the comment that the paper does
not have anything to do with physics.  Today, I would like to clarify
this issue.

In Sec.~\ref{review}, I give a brief review of the subject and explain why
Wigner's paper is essential in understanding modern physics.  In order to
give a more transparent interpretation of his paper, I give a geometrical
interpretation of his work based on the paper which Wigner published with
me in 1987 and 1990~\cite{kiwi87jm,kiwi90jm}.  The purpose of these two
papers was to translate all the earlier works on this subject into a
geometrical language.  The main conclusion of these papers is that the
E(2)-like little group does not share the same geometry as the E(2) group
whose geometry is quite transparent to us.  The geometry of the little
group is that is the cylindrical group dealing with the surface of a
circular cylinder~\cite{boya67}.  The cylindrical axis is parallel to
the momentum.

\begin{figure}

\vspace{25mm}

For a copy of this portrait, send your request to <kim@umdhep.umd.edu>.
Please include your postal address.

\vspace{25mm}

\caption{Eugene Wigner and Albert Einstein. Portrait by Bulent Atalay
(1978).}
\end{figure}

Also shown in these two papers is that the O(3)-like little group,
which can be described in terms of a sphere in the rest frame, becomes
continuously deformed into the symmetry group describing a point moving
on the cylindrical surface as the momentum/mass ratio becomes large.
For the case of electromagnetic four-potential satisfying the Lorentz
condition, the rotation around the axis corresponds to helicity, while
the translation along the direction of the axis corresponds to a gauge
transformation.

In Sec.~\ref{3dim}, we discuss the three-dimensional rotation group and
its contractions to the cylindrical and the two-dimensional Euclidean
groups.  It is shown that both of these contractions can be combined into
a single four-by-four representation.  In Sec.~\ref{cone}, the generators
of the little groups are discussed in the light-cone coordinate system.
It is shown that these generators are identical with the combined
geometry of the cylindrical group and the Euclidean group discussed in
Sec.~\ref{3dim}.  The geometry of Sec.~\ref{3dim} therefore gives a
comprehensive description of the little groups for massive and massless
particles.

\section{Historical Review of Wigner's Little Groups}\label{review}
In 1939, Wigner observed that internal space-time symmetries of
relativistic particles are dictated by their respective little
groups~\cite{wig39}.  The little group is the maximal subgroup of the
Lorentz group which leaves the four-momentum of the particle invariant.
He showed that the little groups for massive and massless particles
are isomorphic to the three-dimensional rotation group and the
two-dimensional Euclidean group respectively.  Wigner's 1939 paper
indeed gives a covariant picture massive particles with spins, and
connects the helicity of massless particle with the rotational degree
of freedom in the group E(2).  This paper also gives many homework
problems for us to solve.

\begin{itemize}
\item First, like the three-dimensional rotation group, E(2) is a
three-parameter group.  It contains two translational degrees of freedom
in addition to the rotation.  What physics is associated with the
translational-like degrees of freedom for the case of the E(2)-like
little group?

\item Second, as is shown by Inonu and Wigner~\cite{inonu53}, the
rotation group O(3) can be contracted to E(2).  Does this mean that the
O(3)-like little group can become the E(2)-like little group in a
certain limit?

\item Third, it is possible to interpret the Dirac equation in terms of
Wigner's representation theory~\cite{barg48}.  Then, why is it not
possible to find a place for Maxwell's equations in the same theory?

\item Fourth, the proton was found to have a finite space-time extension
in 1955~\cite{hofsta55}, and the quark model has been established in
1964~\cite{gell64}.  The concept of relativistic extended particles has
now been firmly established.  Is it then possible to construct a
representation of the Poincar\'e group for particles with space-time
extensions?
\end{itemize}

The list could be endless, but let us concentrate on the above four
questions.  As for the first question, it has been shown by various
authors that the translation-like degrees of freedom in the E(2)-like
little group is the gauge degree of freedom for massless
particles~\cite{janner71}.  As for the second question, it is not
difficult to guess that the O(3)-like little group becomes the E(2)-like
little group in the limit of large momentum/mass~\cite{misra76}.
However, the non-trivial result is that the transverse rotational
degrees of freedom become gauge degrees of freedom~\cite{hks83}.

Then there comes the third question.  Indeed, in 1964~\cite{wein64},
Weinberg found a place for the electromagnetic tensor in Wigner's
representation theory.  He accomplished this by constructing from the
SL(2,c) spinors all the representations of massless fields which are
invariant under the translation-like transformations of the E(2)-like
little group.  Since the translation-like transformations are gauge
transformations, and since the electromagnetic tensor is gauge-invariant,
Weinberg's construction should contain the electric and magnetic fields,
and it indeed does.

Next question is whether it is possible to construct electromagnetic
four-potentials.  After identifying the translation-like degrees of
freedom as gauge degrees of freedom, this becomes a tractable problem.
It is indeed possible to construct gauge-dependent four-potentials
from the SL(2,c) spinors~\cite{hks86}.  Yes, both the field tensor and
four-potential now have their proper places in Wigner's representation
theory.  The Maxwell theory and the Poincar\'e group are perfectly
consistent with each other.

The fourth question is about whether Wigner's little groups are
applicable to high-energy particle physics where accelerators produce
Lorentz-boosted extended hadrons such as high-energy protons.  The
question is whether it is possible to construct a representation of the
Poincar\'e group for hadrons which are believed to be bound states of
quarks~\cite{knp86,fkr71}.  This representation should describe
Lorentz-boosted hadrons.  Next question then is whether those boosted
hadrons give a description of Feynman's parton picture~\cite{fey69}
in the limit of large momentum/mass.  These issues have also been
discussed in the literature~\cite{knp86,kn77a}.

The application of the Poincar\'e group is not limited to relativistic
theories of particles.  This group plays many important roles in classical
mechanics, the theory of superconductivity, as well as in quantum optics.
This new trend makes it more urgent to understand correctly Wigner's papers
on the Lorentz group.  The following sections are based on Wigner's last
papers on this subject~\cite{kiwi87jm,kiwi90jm} where his 1939 paper was
translated into a geometrical language.

\section{Three-dimensional Geometry of the Little Groups}\label{3dim}

The little groups for massive and massless particles are isomorphic
to O(3) and E(2) respectively.  It is not difficult to construct the
O(3)-like geometry of the little group for a massive particle at
rest~\cite{wig39}.  The generators $L_{i}$ of the rotation group satisfy
the commutation relations:
\begin{equation}
[L_{i}, L_{j}] = i\epsilon _{ijk} L_{k} .
\end{equation}
Transformations applicable to the coordinate variables $x, y$, and $z$ are
generated by
\begin{equation}\label{o3gen}
L_{1} = \pmatrix{0&0&0\cr0&0&-i\cr0&i&0} , \quad
L_{2} = \pmatrix{0&0&i\cr0&0&0\cr-i&0&0} , \quad
L_{3} = \pmatrix{0&-i&0\cr i &0&0\cr0&0&0} .
\end{equation}
The Euclidean group E(2) is generated by $L_{3}, P_{1}$ and $P_{2}$, with
\begin{equation}
P_{1} = \pmatrix{0&0&i\cr0&0&0\cr0&0&0} , \qquad
P_{2} = \pmatrix{0&0&0\cr0&0&i\cr0&0&0} ,
\end{equation}
and they satisfy the commutation relations:
\begin{equation}\label{e2com}
[P_{1}, P_{2}] = 0 , \qquad [L_{3}, P_{1}] = iP_{2} ,
\qquad [L_{3}, P_{2}] = -iP_{1} .
\end{equation}
The generator $L_{3}$ is given in Eq.(\ref{o3gen}).  When applied to the
vector space $(x, y, 1)$, $P_{1}$ and $P_{2}$ generate translations on in
the $x y$ plane.  The geometry of E(2) is also quite familiar to us.

Let us transpose the above algebra.  Then $P_{1}$ and $P_{2}$ become
$Q_{1}$ and $Q_{2}$, where
\begin{equation}
Q_{1} = \pmatrix{0&0&0\cr0&0&0\cr i &0&0} , \qquad
Q_{2} = \pmatrix{0&0&0\cr0&0&0\cr0&i&0} ,
\end{equation}
respectively.  Together with $L_{3}$, these generators satisfy the
same set of commutation relations as that for
$L_{3}, P_{1}$, and $P_{2}$ given in Eq.(\ref{e2com})
\begin{equation}
[Q_{1}, Q_{2}] = 0 , \qquad [L_{3}, Q_{1}] = iQ_{2} , \qquad
[L_{3}, Q_{2}] = -iQ_{1} .
\end{equation}
These matrices generate transformations of a point on a circular cylinder.
Rotations around the cylindrical axis are generated by $L_{3}$.  The
$Q_{1}$ and $Q_{2}$ matrices generate the transformation:
\begin{equation}\label{cyltrans}
exp{\left(-i\xi Q_{1} - i\eta Q_{2}\right)} =
\pmatrix{1&0&0\cr0&1&0\cr \xi & \eta & 1} .
\end{equation}
When applied to the space $(x, y, z)$, this matrix changes the value of
$z$ while leaving the $x$ and $y$ variables invariant~\cite{kiwi87jm}.
This corresponds to a translation along the cylindrical axis.  The
$J_{3}$ matrix generates rotations around the axis.  We shall call the
group generated by $J_{3}, Q_{1}$ and $Q_{2}$ the {\em cylindrical group}.

We can achieve the contractions to the Euclidean and cylindrical groups
by taking the large-radius limits of
\begin{equation}
P_{1} = {1\over R} B^{-1} L_{2} B ,
\qquad P_{2} = -{1\over R} B^{-1} L_{1} B ,
\end{equation}
and
\begin{equation}
Q_{1} = -{1\over R}B L_{2}B^{-1} , \qquad
Q_{2} = {1\over R} B L_{1} B^{-1} ,
\end{equation}
where
$$
B(R) = \pmatrix{1&0&0\cr0&1&0\cr0&0&R}  .
$$
The vector spaces to which the above generators are applicable are
$(x, y, z/R)$ and $(x, y, Rz)$ for the Euclidean and cylindrical groups
respectively.  They can be regarded as the north-pole and equatorial-belt
approximations of the spherical surface respectively.

Since $P_{1} (P_{2})$ commutes with $Q_{2} (Q_{1})$, we can consider the
following combination of generators.
\begin{equation}
F_{1} = P_{1} + Q_{1} , \qquad F_{2} = P_{2} + Q_{2} .
\end{equation}
Then these operators also satisfy the commutation relations:
\begin{equation}\label{commuf}
[F_{1}, F_{2}] = 0 , \qquad [L_{3}, F_{1}] = iF_{2} , \qquad
[L_{3}, F_{2}] = -iF_{1} .
\end{equation}
However, we cannot make this addition using the three-by-three matrices
for $P_{i}$ and $Q_{i}$ to construct three-by-three matrices for $F_{1}$
and $F_{2}$, because the vector spaces are different for the $P_{i}$ and
$Q_{i}$ representations.  We can accommodate this difference by creating
two different $z$ coordinates, one with a contracted $z$ and the other
with an expanded $z$, namely $(x, y, Rz, z/R)$.  Then the generators
become four-by-four matrices, and $F_{1}$ and $F_{2}$ take the form
\begin{equation}\label{f1f2}
F_{1} = \pmatrix{0&0&0&i\cr0&0&0&0\cr i &0&0&0\cr0&0&0&0} , \qquad
F_{2} = \pmatrix{0&0&0&0\cr0&0&0&i\cr0&i&0&0\cr0&0&0&0} .
\end{equation}
The rotation generator $L_{3}$ is also a four-by-four matrix:
\begin{equation}\label{2rot}
L_{3} = \pmatrix{0&-i&0&0\cr i&0&0&0\cr0&0&0&0\cr0&0&0&0} .
\end{equation}
These four-by-four matrices satisfy the E(2)-like commutation relations
of Eq.(\ref{commuf}).

Next, let us consider the transformation matrix generated by the above
matrices.  It is easy to visualize the transformations generated by
$P_{i}$ and $Q_{i}$.  It would be easy to visualize the transformation
generated by $F_{1}$ and $F_{2}$, if $P_{i}$ commuted with $Q_{i}$.
However, $P_{i}$ and $Q_{i}$ do not commute with each other, and the
transformation matrix takes a somewhat complicated form:
\begin{equation}\label{compli1}
\exp{\left\{-i(\xi F_{1} + \eta F_{2})\right\}} =
\pmatrix{1&0&0&\xi \cr0 & 1 & 0 & \eta
\cr \xi & \eta & 1 & (\xi ^{2} + \eta ^{2})/2 \cr0&0&0&1}  .
\end{equation}

\section{Little Groups in the Light-cone Coordinate System}\label{cone}
Let us now study the group of Lorentz transformations using the
light-cone coordinate system.  If the space-time coordinate is
specified by $(x, y, z, t)$, then the light-cone coordinate variables are
$(x, y, u, v)$ for a particle moving along the $z$ direction, where
\begin{equation}
u = (z + t)/\sqrt{2} , \qquad v = (z - t)/\sqrt{2} .
\end{equation}
The transformation from the conventional space-time coordinate to the
above system is achieved through a similarity transformation.

It is straight-forward to write the rotation generators $J_{i}$ and boost
generators $K_{i}$ in this light-cone coordinate system~\cite{kiwi90jm}.
If a massive particle is at rest, its little group is generated by $J_{1},
J_{2}$ and $J_{3}$.  For a massless particle moving along the $z$ direction,
the little group is generated by $N_{1}, N_{2}$ and $J_{3}$, where
\begin{equation}
N_{1} = K_{1} - J_{2} , \qquad N_{2} = K_{2} + J_{1} ,
\end{equation}
which can be written in the matrix form as
\begin{equation}
N_{1} = {1\over\sqrt{2}}\pmatrix{0&0&0&i\cr0&0&0&0\cr i &0&0&0\cr0&0&0&0} ,
\qquad N_{2} = {1\over\sqrt{2}}\pmatrix{0&0&0&0\cr0&0&0&i\cr0&i&0&0\cr0&0&0&0}
,
\end{equation}
and $J_{3}$ takes the form of the four-by-four matrix given in Eq.(\ref{2rot})

These matrices satisfy the commutation relations:
\begin{equation}\label{e2comm2}
[J_{3}, N_{1}] =i N_{2} ,\qquad [J_{3}, N_{2}] = -i N_{1} , \qquad
[N_{1}, N_{2}] = 0 .
\end{equation}
Let us go back to $F_{1}$ and $F_{2}$ of Eq.(\ref{f1f2}).  Indeed, they
are proportional to $N_{1}$ and $N_{2}$ respectively.
Since $F_{1}$ and $F_{2}$ are somewhat simpler than $N_{1}$ and $N_{2}$,
and since the commutation relations of Eq.(\ref{e2comm2}) are invariant
under multiplication of $N_{1}$ and $N_{2}$ by constant factors, we shall
hereafter use $F_{1}$ and $F_{2}$ for $N_{1}$ and $N_{2}$.

In the light-cone coordinate system, the boost matrix takes the form
\begin{equation}\label{boost}
B(R) = \exp \pmatrix{-i\rho K_{3}} =
\pmatrix{1&0&0&0\cr0&1&0&0\cr0&0&R&0\cr0&0&0&1/R} ,
\end{equation}
with $\rho = \ln (R)$, and $R = \sqrt{(1 + \beta )/(1 - \beta)}$, where
$\beta$ is the velocity parameter of the particle.  The boost is along the
$z$ direction.  Under this transformation, $x$ and $y$ coordinates are
invariant, and the light-cone variables $u$ and $v$ are transformed as
\begin{equation}
u' = Ru , \qquad v' = v/R .
\end{equation}
If we boost $J_{2}$ and $J_{1}$ and multiply them by $\sqrt{2}/R$, as
$$
W_{1}(R) = -{\sqrt{2} \over R}BJ_{2}B^{-1} = \pmatrix{0&0&-i/R^{2}&i
\cr0&0&0&0 \cr i&0&0&0\cr i/R^{2}&0&0&0}  ,
$$
\begin{equation}\label{w1w2}
W_{2}(R) = {\sqrt{2} \over R} BJ_{1}B^{-1} = \pmatrix{0&0&0&0\cr0&0&-
i/R^{2}&i\cr0&i&0&0\cr0&i/R^{2}&0&0}  ,
\end{equation}
then $W_{1}(R)$ and $W_{2}(R)$ become $F_{1}$ and $F_{2}$ of
Eq.(\ref{f1f2}) respectively in the large-$R$ limit.

The most general form of the transformation matrix is
\begin{equation}
D(\xi, \eta, \alpha ) = D(\xi, \eta, 0)D(0, 0, \alpha) ,
\end{equation}
with
\begin{equation}
D(\xi,\eta,0) = \exp{\left\{-i(\xi F_{1} + \eta F_{2})\right\}} , \qquad
D(0,0,\alpha) = \exp{\left(-i\alpha J_{3}\right)} .
\end{equation}
The matrix $D(0, 0,\alpha)$ represents a rotation around the $z$ axis.
In the light-cone coordinate system, $D(\xi ,\eta ,0)$ takes the form of
Eq.(\ref{compli1}).  It is then possible to decompose it into
\begin{equation}
D(\xi, \eta, 0) = C(\xi, \eta) E(\xi, \eta) S(\xi, \eta) ,
\end{equation}
where
\begin{eqnarray}\label{ces}
S(\xi, \eta) &=& I + {1\over 2} \left[C(\xi, \eta), E(\xi, \eta)\right]
= \pmatrix{1&0&0&0 \cr 0&1&0&0 \cr 0&0&1&(\xi ^{2} + \eta ^{2})/2
\cr 0&0&0&1} ,\nonumber \\[2mm]
E(\xi, \eta) &=& \exp \pmatrix{-i\xi P_{1} - i\eta P_{2}} =
\pmatrix{1&0&0&\xi \cr0&1&0&\eta \cr0&0&1&0\cr0&0&0&1} , \nonumber \\[2mm]
C(\xi ,\eta ) &=& \exp \pmatrix{-i\xi Q_{1} - i\eta Q_{2}} =
\pmatrix{1&0&0&0 \cr 0&1&0&0 \cr \xi & \eta &1&0 \cr0&0&0&1} .
\end{eqnarray}

Let us consider the application of the above transformation matrix to
an electromagnetic four-potential of the form
\begin{equation}
A^{\mu }(x) = A^{\mu } e^{i(kz - \omega t)} ,
\end{equation}
with
\begin{equation}
A^{\mu} = \left(A_{1}, A_{2}, A_{u}, A_{v}\right) ,
\end{equation}
where $A_{u} = (A_{3} + A_{0})/\sqrt{2}$, and $A_{v} = (A_{3} -
A_{0})/\sqrt{2}$.  If we impose the Lorentz condition, the above
four-vector becomes
\begin{equation}
A^{\mu} = \left(A_{1}, A_{2}, A_{u}, 0\right) ,
\end{equation}
The matrix $S(\xi, \eta)$ leaves the above four-vector invariant.
The same is true for the $E(\xi, \eta)$ matrix.  Both $E(\xi, \eta)$
and $S(\xi, \eta)$ become identity matrices when applied to four-vectors
with vanishing fourth component.  Thus only the $C(\xi, \eta)$ matrix
performs non-trivial operations.  As in the case of Eq.(\ref{cyltrans}),
it performs transformations parallel to the cylindrical axis, which in
this case is the direction of the photon momentum.  It leaves the
transverse components of the four vector invariant, but changes the
longitudinal and time-like components at the same rate.  This is a
gauge transformation.

It is remarkable that the algebra of Lorentz transformations given in
this section can be explained in terms of the geometry of deformed
spheres developed in Sec.~\ref{3dim}.

\end{document}